\documentclass{article}
\usepackage{graphicx}
\usepackage{dcolumn}
\usepackage{amsfonts}
\usepackage{latexsym}
\usepackage{amssymb} 
\usepackage{verbatim}
\usepackage{amsthm}
\usepackage{amsmath}

\def\abstract#1{\vskip 7mm 
        \begin{center}{\large Abstract}\par \smallskip
                \begin{minipage}[c]{12cm}
                        \small #1
                \end{minipage}
        \end{center}
}
\def\title#1{\begin{center}{\Large\bf #1}\end{center}}
\def\author#1{\vskip 5mm \begin{center}{#1}\end{center}}
\def\address#1{\begin{center}{\it #1}\end{center}}
\makeatletter
\@ifundefined{lesssim}{}{}
\@ifundefined{gtrsim}{}{}
\def\vereq#1#2{\lower3pt\vbox{\baselineskip1.5pt \lineskip1.5pt
\ialign{$\m@th#1\hfill##\hfil$\crcr#2\crcr\sim\crcr}}}
\makeatother
\begin{document}

\title{%
  Short note on second-order gauge-invariant cosmological perturbation theory 
}
\author{%
  Kouji Nakamura\footnote{E-mail:kouchan@th.nao.ac.jp}
}
\address{%
  Department of Astronomical Science, the Graduate University for
  Advanced Studies, 2-21-1, Osawa, Mitaka, Tokyo 181-8588, Japan
}

\abstract{
  The second order perturbations in Friedmann-Robertson-Walker
  universe filled with a perfect fluid are completely formulated in
  the gauge invariant manner without any gauge fixing.
  All components of the Einstein equations are derived neglecting the
  first order vector and tensor modes.
  These equations imply that the tensor and the vector mode of the
  second order metric perturbations may be generated by the non-linear
  effects of the Einstein equations from the first order density
  perturbations.  
}

Recently, the first order approximation of the early universe
from a homogeneous isotropic one is revealed by the observation
of the CMB by Wilkinson Microwave Anisotropy Probe
(WMAP)\cite{WMAP} and is suggested that fluctuations in the
early universe are adiabatic and Gaussian at least in the first
order approximation.
One of the next theoretical tasks is to clarify the accuracy of
these results, for example, through the non-Gaussianity.
To do this, the second order cosmological perturbation theory
is necessary.


In this article, we show the gauge invariant formulation of the
general relativistic second order cosmological perturbations on
the background Friedmann-Robertson-Walker (FRW) universe 
${\cal M}_{0}$ filled with the perfect fluid whose metric is given by  
\begin{eqnarray}
  g_{ab} = a^{2}(\eta)\left(
    - (d\eta)_{a}(d\eta)_{b}
    + \gamma_{ij}(dx^{i})_{a}(dx^{j})_{b}
  \right),
  \label{eq:background-metric}
\end{eqnarray}
where $\gamma_{ij}$ is the metric on maximally symmetric three
space. The details of our formulation is given in
Refs.\cite{kouchan-gauge-inv}.


The gauge transformation rules for the variable $Q$, which is expanded
as $Q_{\lambda} = Q_{0} + \lambda {}^{(1)}\!Q + \frac{1}{2} \lambda^{2} 
{}^{(2)}\!Q$,
are given by 
\begin{eqnarray}
  \label{eq:Bruni-47-one}
  {}^{(1)}_{\;{\cal Y}}\!Q - {}^{(1)}_{\;{\cal X}}\!Q = 
  {\pounds}_{\xi_{(1)}}Q_{0}, \quad
  {}^{(2)}_{\;\cal Y}\!Q - {}^{(2)}_{\;\cal X}\!Q = 
  2 {\pounds}_{\xi_{(1)}} {}^{(1)}_{\;\cal X}\!Q 
  +\left\{{\pounds}_{\xi_{(2)}}+{\pounds}_{\xi_{(1)}}^{2}\right\} Q_{0},
\end{eqnarray}
where ${\cal X}$ and ${\cal Y}$ represet two different gauge choices,
$\xi_{(1)}^{a}$ and $\xi_{(2)}^{a}$ are generators of the first and
the second order gauge transformations, respectively.


The metric $\bar{g}_{ab}$ on the physical spacetime 
${\cal M}_{\lambda}$ is expanded as   
$\bar{g}_{ab} = g_{ab} + \lambda h_{ab} + \frac{\lambda^{2}}{2} l_{ab}$.
We decompose the components of the first order metric perturbation
$h_{ab}$ into the three sets of variables 
$\{h_{\eta\eta},h_{(VL)},h_{(L)},h_{TL}\}$,
$\{{h_{(V)}}_{i},{h_{(TV)}}_{i}\}$, and
${h_{(TT)ij}}$, which are defined by 
\begin{eqnarray}
  h_{\eta i} &=:& D_{i}h_{(VL)} + h_{(V)i}, \quad
  h_{ij} =: a^{2} h_{(L)} \gamma_{ij} + a^{2}h_{(T)ij}, \quad
  D^{i}h_{(V)i} = 0, \quad
  \gamma^{ij}{h_{(T)}}_{ij} = 0
  , \\ 
  h_{(T)ij} &=:& 
  \left(D_{i}D_{j} - \frac{1}{3}\gamma_{ij}\Delta\right)h_{(TL)}
  + 2 D_{(i}h_{(TV)j)} + {h_{(TT)ij}}
   \quad D^{i} h_{(TV)i} = 0, \quad D^{i} h_{(TT)ij} = 0.
  \label{eq:hij-decomp}
\end{eqnarray}
Inspecting gauge transformation rules (\ref{eq:Bruni-47-one}), we
define a vector field $X_{a}$ by 
\begin{eqnarray}
  X_{a} := X_{\eta}(d\eta)_{a} + X_{i}(dx^{i})_{a}, \quad
  X_{\eta} := h_{(VL)} - \frac{1}{2} a^{2}\partial_{\tau}h_{(TL)} 
  ,
  , \quad
  X_{i} :=  
  a^{2} \left(
      h_{(TV)i}
    + \frac{1}{2} D_{i}h_{(TL)}
  \right)
  ,
\end{eqnarray}
where $X_{a}$ is transformed as
${}_{\;{\cal Y}}\!X_{a}-{}_{\;{\cal X}}\!X_{a} = \xi_{(1)a}$
under the gauge transformation (\ref{eq:Bruni-47-one}).
We can also define the gauge invariant variables for the linear order
metric perturbation by
\begin{eqnarray}
  \label{eq:first-order-gauge-inv-metrc-pert-components}
  {\cal H}_{ab}
  &=& 
  - 2 a^{2} \stackrel{(1)}{\Phi} (d\eta)_{a}(d\eta)_{b}
  + 2 a^{2} \stackrel{(1)}{\nu_{i}} (d\eta)_{(a}(dx^{i})_{b)}
  + a^{2} 
  \left( - 2 \stackrel{(1)}{\Psi} \gamma_{ij} 
    + \stackrel{(1)}{{\chi}_{ij}} \right)
  (dx^{i})_{a}(dx^{j})_{b},
\end{eqnarray}
where
$D^{i}\stackrel{(1)}{\nu_{i}}=\stackrel{(1)}{\chi_{[ij]}}=\stackrel{(1)}{\chi^{i}_{\;i}}=D^{i}\stackrel{(1)}{\chi_{ij}}=0$.
In the cosmological perturbations\cite{Bardeen-1980},
$\{\stackrel{(1)}{\Phi},\stackrel{(1)}{\Psi}\}$,
$\stackrel{(1)}{\nu_{i}}$, $\stackrel{(1)}{\chi_{ij}}$ are
called the scalar, vector, and tensor modes, respectively.
In terms of the variables ${\cal H}_{ab}$ and $X_{a}$, the
original first order metric perturbation $h_{ab}$ is given by 
\begin{eqnarray}
  h_{ab} =: {\cal H}_{ab} + {\pounds}_{X}g_{ab}.
  \label{eq:linear-metric-decomp}
\end{eqnarray}
Since the scalar mode dominates in the early universe, we assume
that $\stackrel{(1)}{\nu_{i}}=\stackrel{(1)}{\chi_{ij}}=0$ in
this article.


As shown in Ref.\cite{kouchan-gauge-inv}, through the above variables
$X_{a}$ and $h_{ab}$, the second order metric perturbation $l_{ab}$ is
decomposed as 
\begin{eqnarray}
  \label{eq:H-ab-in-gauge-X-def-second-1}
  l_{ab}
  =:
  {\cal L}_{ab} + 2 {\pounds}_{X} h_{ab}
  + \left(
      {\pounds}_{Y}
    - {\pounds}_{X}^{2} 
  \right)
  g_{ab}.
\end{eqnarray}
The variables ${\cal L}_{ab}$ and $Y^{a}$ are the gauge
invariant and variant parts of $l_{ab}$, respectively.
The vector field $Y_{a}$ is transformed as 
${}_{\;{\cal Y}}\!Y_{a}-{}_{\;{\cal X}}\!Y_{a} = \xi_{(2)}^{a} +
[\xi_{(1)},X]^{a}$ under the gauge transformations
(\ref{eq:Bruni-47-one}).
The components of ${\cal L}_{ab}$ are given by 
\begin{eqnarray}
  \label{eq:second-order-gauge-inv-metrc-pert-components}
  {\cal L}_{ab}
  &=& 
  - 2 a^{2} \stackrel{(2)}{\Phi} (d\eta)_{a}(d\eta)_{b}
  + 2 a^{2} \stackrel{(2)}{\nu_{i}} (d\eta)_{(a}(dx^{i})_{b)}
  + a^{2} 
  \left( - 2 \stackrel{(2)}{\Psi} \gamma_{ij} 
    + \stackrel{(2)}{{\chi}_{ij}} \right)
  (dx^{i})_{a}(dx^{j})_{b},
\end{eqnarray}
where
$D^{i}\stackrel{(2)}{\nu_{i}}=\stackrel{(2)}{\chi_{[ij]}}=\stackrel{(2)}{\chi^{i}_{\;\;i}}=D^{i}\stackrel{(2)}{\chi_{ij}}
= 0$.


As shown in Ref.\cite{kouchan-gauge-inv}, by using the above variables
$X_{a}$ and $Y_{a}$, we can find the gauge invariant variables for the
perturbations of an arbitrary field as 
\begin{eqnarray}
  \label{eq:matter-gauge-inv-def-1.0}
  {}^{(1)}\!{\cal Q} := {}^{(1)}\!Q - {\pounds}_{X}Q_{0},
  , \quad 
  {}^{(2)}\!{\cal Q} := {}^{(2)}\!Q - 2 {\pounds}_{X} {}^{(1)}Q 
  - \left\{ {\pounds}_{Y} - {\pounds}_{X}^{2} \right\} Q_{0}.
\end{eqnarray}
As the matter contents, in this article, we consider the perfect
fluid whose energy-momentum tensor is given by 
$\bar{T}_{a}^{\;\;b} = \left(\bar{\epsilon}+\bar{p}\right)
\bar{u}_{a}\bar{u}^{b} + \bar{p}\delta_{a}^{\;\;b}$.
We expand these fluid components $\bar{\epsilon}$, $\bar{p}$,
and $\bar{u}_{a}$ as
\begin{eqnarray}
  \bar{\epsilon}
  =
  \epsilon
  + \lambda \stackrel{(1)}{\epsilon}
  + \frac{1}{2} \lambda^{2} \stackrel{(2)}{\epsilon}
  , 
  \quad
  \bar{p}
  =
  p
  + \lambda \stackrel{(1)}{p}
  + \frac{1}{2} \lambda^{2} \stackrel{(2)}{p} 
  ,
  \quad
  \bar{u}_{a}
  =
  u_{a}
  + \lambda \stackrel{(1)}{u}_{a}
  + \frac{1}{2} \lambda^{2} \stackrel{(2)}{u}_{a}p 
  .
  \label{eq:Taylor-expansion-of-four-velocity}
\end{eqnarray}
Following the definitions (\ref{eq:matter-gauge-inv-def-1.0}), we
easily obtain the corresponding gauge invariant variables for these
perturbations of the fluid components:
\begin{eqnarray}
  \stackrel{(1)}{{\cal E}} 
  &:=& \stackrel{(1)}{\epsilon} - {\pounds}_{X}\epsilon, \quad
  \stackrel{(1)}{{\cal P}}
  := \stackrel{(1)}{p} - {\pounds}_{X}p, \quad
  \stackrel{(1)}{{\cal U}_{a}}
  := \stackrel{(1)}{(u_{a})} - {\pounds}_{X}u_{a}, \nonumber
 \quad
  \stackrel{(2)}{{\cal E}} 
  := \stackrel{(2)}{\epsilon} 
  - 2 {\pounds}_{X} \stackrel{(1)}{\epsilon}
  - \left\{
    {\pounds}_{Y}
    -{\pounds}_{X}^{2}
  \right\} \epsilon
  , \nonumber
  \\
  \stackrel{(2)}{{\cal P}}
  &:=& \stackrel{(2)}{p}
  - 2 {\pounds}_{X} \stackrel{(1)}{p}
  - \left\{
    {\pounds}_{Y}
    -{\pounds}_{X}^{2}
  \right\} p
  , 
  \quad
  \label{eq:kouchan-016.18}
  \stackrel{(2)}{{\cal U}_{a}}
  := \stackrel{(2)}{(u_{a})}
  - 2 {\pounds}_{X} \stackrel{(1)}{u_{a}}
  - \left\{
    {\pounds}_{Y}
    -{\pounds}_{X}^{2}
  \right\} u_{a}.
  \nonumber
\end{eqnarray}
Through $\bar{g}^{ab}\bar{u}_{a}\bar{u}_{b}=g^{ab}u_{a}u_{b}=-1$
and neglecting the rotational part in
$\stackrel{(1)}{{\cal U}_{a}}$, the components of
$\stackrel{(1)}{{\cal U}_{a}}$ are given by 
$\stackrel{(1)}{{\cal U}_{a}} = - a \stackrel{(1)}{\Phi} (d\eta)_{a}
+ a D_{i} \stackrel{(1)}{v} (dx^{i})_{a}$.


We also expand the Einstein tensor as 
$\bar{G}_{a}^{\;\;b} = G_{a}^{\;\;b} + \lambda {}^{(1)}\!G_{a}^{\;\;b}
+ \frac{1}{2} \lambda^{2} {}^{(2)}\!G_{a}^{\;\;b}$.
From Eqs.(\ref{eq:linear-metric-decomp}) and
(\ref{eq:H-ab-in-gauge-X-def-second-1}), each order
perturbation of the Einstein tensor is given by 
\begin{eqnarray}
  {}^{(1)}\!G_{a}^{\;\;b}
  =
  {}^{(1)}{\cal G}_{a}^{\;\;b}\left[{\cal H}\right]
  + {\pounds}_{X}G_{a}^{\;\;b}
  ,
  \quad
  {}^{(2)}\!G_{a}^{\;\;b}
  =
  {}^{(1)}{\cal G}_{a}^{\;\;b}\left[{\cal L}\right]
  + {}^{(2)}{\cal G}_{a}^{\;\;b}\left[{\cal H}, {\cal H}\right]
  + 2 {\pounds}_{X} {}^{(1)}\!G_{a}^{\;\;b}
  + \left\{
    {\pounds}_{Y}
    -{\pounds}_{X}^{2}
  \right\} G_{a}^{\;\;b}
\end{eqnarray}
as expected from Eqs.~(\ref{eq:matter-gauge-inv-def-1.0}).
Here, ${}^{(1)}{\cal G}_{a}^{\;\;b}\left[{\cal H}\right]$ and 
${}^{(1)}{\cal G}_{a}^{\;\;b}\left[{\cal L}\right]
+ {}^{(2)}{\cal G}_{a}^{\;\;b}\left[{\cal H}, {\cal H}\right]$
are gauge invariant parts of the frist and the second order
perturbations of the Einstein tensor, respectively.
On the other hand, the energy mometum tensor of the perfect fluid is
also expanded as 
$\bar{T}_{a}^{\;\;b} = T_{a}^{\;\;b} + \lambda {}^{(1)}\!T_{a}^{\;\;b}
+ \frac{1}{2} \lambda^{2} {}^{(2)}\!T_{a}^{\;\;b}$
and ${}^{(1)}\!T_{a}^{\;\;b}$ and ${}^{(2)}\!T_{a}^{\;\;b}$ are
also given in the form 
\begin{eqnarray}
  {}^{(1)}\!T_{a}^{\;\;b}
  =
  {}^{(1)}\!{\cal T}_{a}^{\;\;b}
  + {\pounds}_{X}T_{a}^{\;\;b}
  ,
  \quad
  {}^{(2)}\!T_{a}^{\;\;b}
  =
  {}^{(2)}\!{\cal T}_{a}^{\;\;b}
  + 2 {\pounds}_{X} {}^{(1)}\!T_{a}^{\;\;b}
  + \left\{
    {\pounds}_{Y}
    -{\pounds}_{X}^{2}
  \right\} T_{a}^{\;\;b}
\end{eqnarray}
through the definitions (\ref{eq:kouchan-016.18}) of the
gauge invariant variables of the fluid components.
Here, ${}^{(1)}\!{\cal T}_{a}^{\;\;b}$ and 
${}^{(2)}\!{\cal T}_{a}^{\;\;b}$ are gauge invariant part of the  
first and the second order perturbation of the energy momentum
tensor, respectively. 
Then, the first and the second order perturbations of the
Einstein equation are necessarily given in term of gauge
invaraint variables:
\begin{eqnarray}
  \label{eq:linear-order-Einstein-equation}
  {}^{(1)}{\cal G}_{a}^{\;\;b}\left[{\cal H}\right]
  =
  8\pi G {}^{(1)}{\cal T}_{a}^{\;\;b},
  \quad
  {}^{(1)}{\cal G}_{a}^{\;\;b}\left[{\cal L}\right]
  + {}^{(2)}{\cal G}_{a}^{\;\;b}\left[{\cal H}, {\cal H}\right]
  =
  8\pi G \;\; {}^{(2)}{\cal T}_{a}^{\;\;b}. 
\end{eqnarray}


The traceless scalar part of the spatial component of
the first equation in Eq.(\ref{eq:linear-order-Einstein-equation})
yields $\stackrel{(1)}{\Psi} = \stackrel{(1)}{\Phi}$, and the other
components of Eq.~(\ref{eq:linear-order-Einstein-equation})
give well-known equations\cite{Bardeen-1980}.


Since we neglect the first order vector and tensor modes,
$\stackrel{(2)}{{\cal U}_{a}}$ is given by 
\begin{eqnarray}
  \stackrel{(2)}{{\cal U}_{a}}
  &=&
  a \left(
    \left(\stackrel{(1)}{\Phi}\right)^{2}
    - D_{i}\stackrel{(1)}{v} D^{i}\stackrel{(1)}{v}
    - \stackrel{(2)}{\Phi}
  \right) (d\eta)_{a}
  +
  a \left(
    D_{i} \stackrel{(2)}{v}
    +
    \stackrel{(2)}{{\cal V}_{i}}
  \right) (dx^{i})_{a},
\end{eqnarray}
where $D^{i} \stackrel{(2)}{{\cal V}_{i}} = 0$.
All components of the second equation in
Eq.~(\ref{eq:linear-order-Einstein-equation}) are summarized as
follows: 
As the scalar parts, we have 
\begin{eqnarray}
  4\pi G a^{2} \stackrel{(2)}{{\cal E}}
  &=&
  \left(
    - 3 {\cal H} \partial_{\eta}
    +   \Delta
    + 3 K
    - 3 {\cal H}^{2}
  \right)
  \stackrel{(2)}{\Phi}
  - \Gamma_{0}
  +
  \frac{3}{2}
  \left(
    \Delta^{-1} D^{i}D_{j}\Gamma_{i}^{\;\;j}
    - \frac{1}{3} \Gamma_{k}^{\;\;k}
  \right)
  \nonumber\\
  && \quad
  -
  \frac{9}{2}
  {\cal H} \partial_{\eta}
  \left( \Delta + 3 K \right)^{-1}
  \left(
    \Delta^{-1} D^{i}D_{j}\Gamma_{i}^{\;\;j}
    - \frac{1}{3} \Gamma_{k}^{\;\;k}
  \right)
  \label{eq:kouchan-18.79}
  , \\
  8\pi G a^{2} (\epsilon + p) D_{i}\stackrel{(2)}{v} 
  &=&
  - 2 \partial_{\eta}D_{i}\stackrel{(2)}{\Phi}
  - 2 {\cal H} D_{i}\stackrel{(2)}{\Phi}
  + D_{i} \Delta^{-1} D^{k}\Gamma_{k}
  \nonumber\\
  && \quad
  - 3 \partial_{\eta}D_{i}
  \left( \Delta + 3 K \right)^{-1}
  \left(
    \Delta^{-1} D^{i}D_{j}\Gamma_{i}^{\;\;j}
    - \frac{1}{3} \Gamma_{k}^{\;\;k}
  \right)
  \label{eq:second-velocity-scalar-part-Einstein}
  , \\
  4 \pi G a^{2} \stackrel{(2)}{{\cal P}}
  &=&
  \left(
      \partial_{\eta}^{2} 
    + 3{\cal H} \partial_{\eta}
    - K
    + 2\partial_{\eta}{\cal H}
    + {\cal H}^{2}
  \right)
  \stackrel{(2)}{\Phi}
  -
  \frac{1}{2}
  \Delta^{-1} D^{i}D_{j}\Gamma_{i}^{\;\;j}
  \nonumber\\
  && \quad
  +
  \frac{3}{2}
  \left(
        \partial_{\eta}^{2} 
    + 2 {\cal H} \partial_{\eta}
  \right)
  \left( \Delta + 3 K \right)^{-1}
  \left(
    \Delta^{-1} D^{i}D_{j}\Gamma_{i}^{\;\;j}
    - \frac{1}{3} \Gamma_{k}^{\;\;k}
  \right)
  ,\\
  \label{eq:kouchan-18.80}
  \stackrel{(2)}{\Psi} - \stackrel{(2)}{\Phi}
  &=&
  \frac{3}{2}
  \left( \Delta + 3 K \right)^{-1}
  \left(
    \Delta^{-1} D^{i}D_{j}\Gamma_{i}^{\;\;j}
    - \frac{1}{3} \Gamma_{k}^{\;\;k}
  \right)
  .
  \label{eq:kouchan-18.65}
\end{eqnarray}
where ${\cal H} = \partial_{\eta}a/a$ and 
\begin{eqnarray}
  \Gamma_{0}
  &:=&
  8\pi G a^{2} (\epsilon + p) D^{i}\stackrel{(1)}{v} D_{i}\stackrel{(1)}{v} 
  - 3 D_{k}\stackrel{(1)}{\Phi} D^{k}\stackrel{(1)}{\Phi}
  - 3 \left(\partial_{\eta}\stackrel{(1)}{\Phi}\right)^{2}
  -  8 \stackrel{(1)}{\Phi} \Delta \stackrel{(1)}{\Phi}
  - 12 \left( K + {\cal H}^{2} \right) \left(\stackrel{(1)}{\Phi}\right)^{2}
  ,
  \nonumber\\
  \Gamma_{i}
  &:=&
  - 16 \pi G a^{2} \left(
     \stackrel{(1)}{{\cal E}} + \stackrel{(1)}{{\cal P}}
   \right)D_{i}\stackrel{(1)}{v}
  +         12  {\cal H} \stackrel{(1)}{\Phi} D_{i}\stackrel{(1)}{\Phi}
  -          4  \stackrel{(1)}{\Phi} \partial_{\eta}D_{i}\stackrel{(1)}{\Phi}
  -          4  \partial_{\eta}\stackrel{(1)}{\Phi} D_{i}\stackrel{(1)}{\Phi}
  ,
  \label{eq:kouchan-18.56}
  \\
  \Gamma_{i}^{\;\;j}
  &:=&
    16 \pi G a^{2} (\epsilon+p) D_{i}\stackrel{(1)}{v} D^{j}\stackrel{(1)}{v}
  -  4  D_{i}\stackrel{(1)}{\Phi} D^{j}\stackrel{(1)}{\Phi}
  -  8  \stackrel{(1)}{\Phi} D_{i}D^{j}\stackrel{(1)}{\Phi}
  \nonumber\\
  && 
  + 2 \left(
      3 D_{k}\stackrel{(1)}{\Phi} D^{k}\stackrel{(1)}{\Phi}
    + 4 \stackrel{(1)}{\Phi} \Delta \stackrel{(1)}{\Phi}
    +   \left(\partial_{\eta}\stackrel{(1)}{\Phi}\right)^{2}
    + 4 \left(
      2 \partial_{\eta}{\cal H} + K + {\cal H}^{2}
    \right) \left(\stackrel{(1)}{\Phi}\right)^{2}
    + 8 {\cal H} \stackrel{(1)}{\Phi} \partial_{\eta}\stackrel{(1)}{\Phi}
  \right)
  \gamma_{i}^{\;\;j}
  .
  \nonumber
\end{eqnarray}
As the vector parts, we have
\begin{eqnarray}
  8\pi G a^{2} (\epsilon + p) \stackrel{(2)}{{\cal V}}_{i}
  &=&
  \frac{1}{2} \left(
    \Delta
    + 2 K
  \right)
  \stackrel{(2)}{\nu_{i}}
  +\left(
    \Gamma_{i}
    -
    D_{i} \Delta^{-1} D^{k} \Gamma_{k}
  \right)
  \label{eq:kouchan-18.59-vorticity}
  , \\
  \partial_{\eta} \left(a^{2}\stackrel{(2)}{\nu_{i}}\right)
  &=&
  2a^{2}
  \left( \Delta + 2 K \right)^{-1}
  \left\{
    D_{i} \Delta^{-1} D^{k}D_{l}\Gamma_{k}^{\;\;l}
    - D_{k}\Gamma_{i}^{\;\;k}
  \right\}
  .
  \label{eq:kouchan-18.66}
\end{eqnarray}
As the tensor parts, we have the evolution equation of
$\stackrel{(2)\;\;\;\;}{\chi_{ij}}$
\begin{eqnarray}
  &&
  \left(
    \partial_{\eta}^{2} + 2 {\cal H} \partial_{\eta} + 2 K  - \Delta
  \right)
  \stackrel{(2)\;\;\;\;}{\chi_{ij}}
  \nonumber\\
  &=&
  2 \Gamma_{ij}
  - \frac{2}{3} \gamma_{ij} \Gamma_{k}^{\;\;k}
  - 3
  \left(
    D_{i}D_{j} - \frac{1}{3} \gamma_{ij} \Delta
  \right) 
  \left( \Delta + 3 K \right)^{-1}
  \left(
    \Delta^{-1} D^{k}D_{l}\Gamma_{k}^{\;\;l}
    - \frac{1}{3} \Gamma_{k}^{\;\;k}
  \right)
  \nonumber\\
  &&
  + 4
  \left( 
    D_{(i}\left( \Delta + 2 K \right)^{-1}
    D_{j)}\Delta^{-1}D^{l}D_{k}\Gamma_{l}^{\;\;k}
    - D_{(i} \left( \Delta + 2 K \right)^{-1} D^{k}\Gamma_{j)k}
  \right)
  ,
  \label{eq:kouchan-18.68}
\end{eqnarray}
Equations (\ref{eq:kouchan-18.66}) and (\ref{eq:kouchan-18.68})
imply that the second order vector and tensor modes may be
generated due to the scalar-scalar mode coupling of the first
order perturbation.


Further, the equations (\ref{eq:kouchan-18.79}) and
(\ref{eq:kouchan-18.80}) are reduced to the single equation for
$\stackrel{(2)}{\Phi}$ 
\begin{eqnarray}
  && 
  \left(
      \partial_{\eta}^{2} 
    + 3 {\cal H} (1 + c_{s}^{2}) \partial_{\eta}
    -   c_{s}^{2} \Delta
    + 2 \partial_{\eta}{\cal H}
    + (1 + 3 c_{s}^{2}) ({\cal H}^{2} - K)
  \right)
  \stackrel{(2)}{\Phi}
  \nonumber\\
  &=&
  4\pi G a^{2} \left\{
    \tau \stackrel{(2)}{{\cal S}}
    + \frac{\partial c_{s}^{2}}{\partial\epsilon}
    \left(\stackrel{(1)}{{\cal E}}\right)^{2}
    + 2 \frac{\partial c_{s}^{2}}{\partial S}
    \stackrel{(1)}{{\cal E}}
    \stackrel{(1)}{{\cal S}}
    + \frac{\partial\tau}{\partial S}
    \left(\stackrel{(1)}{{\cal S}}\right)^{2}
  \right\}
  +
  \frac{3}{2}
  \left(c_{s}^{2} + \frac{1}{3}\right)
  \left(
    \Delta^{-1} D^{i}D_{j}\Gamma_{i}^{\;\;j}
    - \frac{1}{3} \Gamma_{k}^{\;\;k}
  \right)
  \nonumber\\
  && \quad
  - c_{s}^{2} \Gamma_{0}
  + \frac{1}{6} \Gamma_{k}^{\;\;k}
  -
  \frac{3}{2}
  \left(
      \partial_{\eta}^{2} 
    + \left( 2 + 3 c_{s}^{2} \right) {\cal H} \partial_{\eta}
  \right)
  \left( \Delta + 3 K \right)^{-1}
  \left(
    \Delta^{-1} D^{i}D_{j}\Gamma_{i}^{\;\;j}
    - \frac{1}{3} \Gamma_{k}^{\;\;k}
  \right)
  .
  \label{eq:second-order-Einstein-scalar-master-eq}
\end{eqnarray}
Here, we have used the second order perturbation of the equation
of state for the fluid components
\begin{eqnarray}
  \stackrel{(2)}{{\cal P}}
  =
  c_{s}^{2}
  \stackrel{(2)}{{\cal E}} 
  + 
  \tau
  \stackrel{(2)}{{\cal S}} 
  +
  \frac{\partial c_{s}^{2}}{\partial\epsilon}
  \stackrel{(1)}{{\cal E}}^{2}
  + 2
  \frac{\partial c_{s}^{2}}{\partial S}
  \stackrel{(1)}{{\cal E}}
  \stackrel{(1)}{{\cal S}}
  +
  \frac{\partial\tau}{\partial S}
  \stackrel{(1)}{{\cal S}}^{2},
  \label{eq:second-order-equation-of-state-gauge-inv}
\end{eqnarray}
where $\stackrel{(1)}{{\cal S}}$ and $\stackrel{(2)}{{\cal S}}$
are the gauge invariant entropy perturbation of the first and
second order, respectively, we denoted that $c_{s}^{2} :=
\partial\bar{p}/\partial\bar{\epsilon}$ and $\tau :=
\partial\bar{p}/\partial\bar{S}$.
The equation (\ref{eq:second-order-Einstein-scalar-master-eq})
will be useful to discus non-linear effects in the CMB
physics\cite{Non-Gaussianity-in-CMB}.
We also derive the similar equations in the case where the
matter content of the universe is a single scalar
field\cite{kouchan-gauge-inv}.


Now, we are developing our formulation to the case in which the first
order vector and tensor modes are not negligible. 
In some inflationary scenario, the tensor mode are also generated by
the quantum fluctuations.
This extension is necessary to clarify the evolution of the second
order perturbation in the existence of the first order tensor mode.
Further, to apply this formulation to clarify the non-linear effects
in CMB physics\cite{Non-Gaussianity-in-CMB}, we have to extend our
formulation to multi-field system and the Einstein Boltzmann system.
This extension of our formulation is one of our future works.


Moreover, the rotational part of the fluid velocity in
Eqs.~(\ref{eq:kouchan-18.59-vorticity}) of the vector mode is
also important in the early universe because this part of the
fluid velocity is related to the generation of the magnetic
field in the early universe\cite{Matarrese-etal-2005}.
The generation of the tensor mode by
Eq.~(\ref{eq:kouchan-18.68}) is also interesting, since this is
one of the generation process of gravitational waves.
We have already known that the fluctuations of the scalar mode
exist in the early universe from the anisotropy of the
CMB\cite{Mollerach-Harari-Matarrse-2004}.
Hence, the generation of the vector mode and tensor mode due to
the second order perturbation will give the lower limit of these
modes in the early universe.


\end{document}